\documentclass[aps,prc,preprint,groupedaddress]{revtex4}

\usepackage{graphicx}

\usepackage{amssymb}

\begin{document}




\title{Search for Oscillation of the Electron-Capture Decay Probability of $^{142}$Pm}


\author{P.A.~Vetter}
\email{pavetter@lbl.gov}
\affiliation{Lawrence Berkeley National Laboratory, One Cyclotron Road, Berkeley, California 94720}

\author{R.M.~Clark}
\affiliation{Lawrence Berkeley National Laboratory, One Cyclotron Road, Berkeley, California 94720}

\author{J.~Dvorak}
\affiliation{Lawrence Berkeley National Laboratory, One Cyclotron Road, Berkeley, California 94720}

\author{S.J.~Freedman}
\affiliation{Department of Physics, University of California at Berkeley and
Lawrence Berkeley National Laboratory, Berkeley, California 94720}

\author{K.E.~Gregorich}
\affiliation{Lawrence Berkeley National Laboratory, One Cyclotron Road, Berkeley, California 94720}

\author{H.B.~Jeppesen}
\affiliation{Lawrence Berkeley National Laboratory, One Cyclotron Road, Berkeley, California 94720}

\author{D.~Mittelberger}
\affiliation{Department of Physics, University of California at Berkeley and
Lawrence Berkeley National Laboratory, Berkeley, California 94720}

\author{M.~Wiedeking}
\affiliation{Lawrence Livermore National Laboratory, Livermore, California 94551}

\date{\today}

\begin{abstract}
We have searched for time modulation of the electron capture decay probability of $^{142}$Pm in an attempt to confirm a recent claim from a group at the Gesellschaft f\"{u}r Schwerionenforschung (GSI). We produced $^{142}$Pm via the $^{124}$Sn($^{23}$Na, 5n)$^{142}$Pm reaction at the Berkeley 88-Inch Cyclotron with a bombardment time short compared to the reported modulation period. Isotope selection by the Berkeley Gas-filled Separator is followed by implantation and a long period of monitoring the $^{142}$Nd K$_{\alpha}$ x-rays from the daughter. The decay time spectrum of the x-rays is well-described by a simple exponential and the measured half-life of 40.68(53)~seconds is consistent with the accepted value. We observed no oscillatory modulation at the proposed frequency at a level 31 times smaller than that reported by Litvinov {\it et al.} (Phys.~Lett.~B 664 (2008) 162).  A literature search for previous experiments that might have been sensitive to the reported modulation uncovered another example in $^{142}$Eu electron-capture decay.  A reanalysis of the published data shows no oscillatory behavior.

\end{abstract}

\maketitle


A recent paper reported observation of an oscillating decay rate of the isotopes $^{140}$Pr and $^{142}$Pm when electron capture decays were measured using highly charged ions with a Schottky mass spectroscopy technique \cite{litvinov08}.  The authors concluded that the decay activity oscillated in time with a period of about 7 seconds and a relative amplitude of 20\% for both isotopes, and attributed the oscillating behavior to interference between neutrino mass eigenstates in the two-body kinematics for electron capture decay of the hydrogen-like ions.  If confirmed, this effect is surprising and might offer a new avenue for studying neutrino mixing.  References \cite{faber08,ivanov08} suggest quantitative explanations for the observations of Ref.\ \cite{litvinov08} in terms of two-species neutrino mixing.  Other authors argue strongly in Refs.\ \cite{giunti08a,giunti08b} that associating the claimed decay rate modulation with neutrino oscillation is inconsistent with well-established principles of quantum mechanics.  We do not wish to enter the debate about the consistency of the experimental claims with the theoretical treatment of neutrino oscillations and quantum mechanics in the present paper.  Instead we focus on the experimental issues and describe our attempt to confirm the findings of the GSI group.

The experiment at GSI exploited several unique features of the heavy ion synchrotron and storage ring facility \cite{geissel92}.  The measurement was made on hydrogen-like ions of $^{142}$Pm and $^{140}$Pr produced in flight and then mass separated.  After a short period of stochastic cooling, one or two ions at a time are captured, and the cyclotron frequency is measured as the ions coast in the ring.  The relevant parameter is the time to electron capture following ion injection.  With this elegant technique, the daughter ions (which have the same charge but slightly different masses) are also trapped and the time of electron capture is determined as a discontinuous change of the cyclotron frequency.  The authors find that their decay data are well-described by 
\begin{eqnarray}
\label{eq:oscdec}
\frac{dN_{EC}\left(t\right) }{dt} & = &\Gamma\left(t\right) =  B + N_{0}\cdot e^{-t \ln2 /T_{1/2}} \nonumber \\ 
 & & \times \left[ 1 + A\cdot \cos\left( \omega t + \phi \right) \right] .
\end{eqnarray}
According to the model in Refs.\ \cite{faber08,ivanov08,lipkin08a}, the oscillation period is $T_{d} = 2\pi /\omega \propto \frac{\gamma M_{d}}{\Delta m_{21}^{2}}$, where $M_{d}$ is the mass of the daughter nucleus, $\gamma$ is the Lorentz factor of the moving stored ions ($\gamma = 1.43$ for the stored ions in the GSI experiments), and $\Delta m_{21}^{2}$ is the squared mass difference between the two participating neutrino mass eigenstates.

We have considered the possibility that the oscillating decay rate effect might have been missed in previous experiments studying electron capture or even ordinary beta decay using more traditional radiation detectors and implanted sources of neutral atoms.  We searched the literature for decays with relatively high electron capture probability and with half-lives within a range suggested by the model in \cite{ivanov08} ($T_{d}\propto \frac{M_{d}}{\Delta m_{21}^{2} }$). The modulation period depends linearly upon the mass of the decaying nucleus, and in nuclei around A=150, the period (at rest in the lab frame) is about 5 seconds.  There are numerous examples of measured decays in this mass region that have characteristics similar to $^{142}$Pm and $^{140}$Pr.  In most experiments to measure decay properties, a typical procedure is to bombard the target for a time comparable to the lifetime of the isotope of interest.  This timing optimizes the data collection rate, but reduces sensitivity to a time-modulated decay rate because the modulation for nuclei produced at different times would destructively interfere.  
An oscillating effect would be most clearly seen when decay time and modulation period are very similar.  On the other hand, the effect would be difficult to see for lifetimes much shorter than the modulation period.  For $^{140}$Pr and $^{142}$Pm, typical long bombardment and production times would preclude the observation of a 5 to 10 second modulated decay time spectrum.  

We found one earlier experiment which did have the necessary short time bombardment followed by a long counting period:  a study of the isomeric decay of $^{142}$Eu to states in $^{142}$Sm \cite{kennedy75}.  In that work, the $^{142}$Eu was prepared using short irradiations from 0.1 to 2.0~s, which would preserve a 5 to 10 second oscillation, but with a diminution of the modulation amplitude from the suggested 20\% to between 19.9\% and 15\% (depending on the irradiation time).  The decays of the ground $(1^{+})$ ($T_{1/2} = 2.4$ s) and $(8^{-})$ isomer ($T_{1/2} = 1.22$ m) states were observed by subsequent gamma decay  of excited states of $^{142}$Sm populated uniquely by ground or isomer decays.  Figure 1 from Ref.~\cite{kennedy75} shows the decay of the gamma activity produced by decays of $^{142}$Eu$^{m}$, with 1 second time bins, extending to 50 seconds after bombardment.  The electron capture probability for the 1.22~m $(8^{-})$ isomer decay is 17\%.

\begin{figure}
\begin{center}
\includegraphics[width=90mm,keepaspectratio=true]{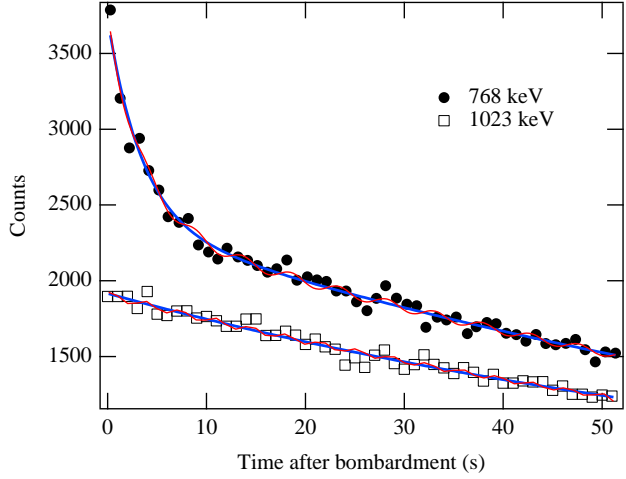}
\caption{Data from Figure 1 in Ref.\ \cite{kennedy75} analyzed to search for decay oscillations.  The vertical axis represents measured counts in the photopeak for two transitions in $^{142}$Sm fed by ground and isomer state decays of $^{142}$Eu.  The minimum chi-squared fits with Eq.\ \ref{eq:oscdec} (red) and exponential decay (blue) are shown with the data.  The 768 keV data shows decay of both the 2.4 s ground state and the 1.22 m ($8^{+}$) state of $^{142}$Eu.}
\label{fig:142Eudatfit}
\end{center}
\end{figure}

\begin{figure}
\begin{center}
\includegraphics[width=90mm,keepaspectratio=true]{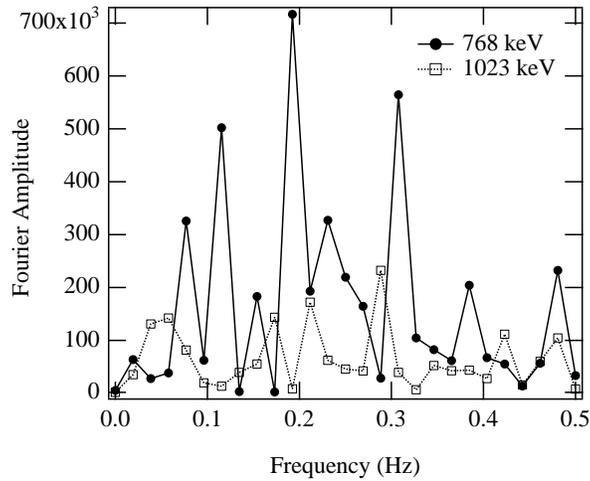}
\caption{A Fourier transform of data from Fig.\ \ref{fig:142Eudatfit} when the data are fit with just exponential decay, without an oscillating term, assuming the known lifetimes of the ground and isomeric states in $^{142}$Eu.  The Fourier transform is taken of residuals (data minus fit) to search for a frequency peak not accounted for by an exponential decay term. No peak is resolved at either 0.14 Hz (7 seconds) or 0.2 Hz (5 seconds) in either data set.}
\label{fig:142EuFFT}
\end{center}
\end{figure}

The data from Figure 1 of Ref.~\cite{kennedy75} show no obvious oscillations.  A fit of that data (shown in this paper in Fig.\ \ref{fig:142Eudatfit}) using Eq.\ \ref{eq:oscdec} finds a minimum of $\chi^{2}/\mathrm{dof} = 0.792$ with an oscillation period of 3.540(53)~seconds, an amplitude of $A = 0.0136(55)$, and a phase $\phi = 5.6(8)$~rad for the 1023~keV data.  For the 768~keV data, we find a minimum $\chi^{2}/\mathrm{dof} = 1.36$ with an oscillation period of 4.854(73)~seconds, $A = 0.0133(84)$, and $\phi = 1.13(55)$~rad.  When performing fits to the data using Eq.\ \ref{eq:oscdec} we restrict the phase to $0<\phi<2\pi$ and ensure $A>0$.  Fits to simple exponential decay find $\chi^{2}/\mathrm{dof} = 0.868$ for the 1023~keV data and $\chi^{2}/\mathrm{dof} = 1.48$ for the 768~keV data.  The Fourier transform power spectra in Fig.\ \ref{fig:142EuFFT} are calculated for the residuals to simple exponential fits.  The Fourier spectra show no well-resolved peaks corresponding to 5 second oscillation (or 7 second oscillation, under the assumption that the oscillation period is not proportional to the Lorentz factor $\gamma$). Our $^{142}$Eu analysis would limit an oscillatory term to be a factor of about 3 smaller than that reported in Ref.\ \cite{litvinov08}, if we assume that our uncertainty in $A$ ($\pm0.0084$) represents our sensitivity to an oscillating term, and if we correct for the electron capture branching ratio and account for the small reduction in an oscillation amplitude from the (up to) 2~s bombardment time.  Our best-fit period and phase parameters disagree strongly with the results of \cite{litvinov08}.  The electron capture branching ratio also depends on the ionization state of the parent atom, as shown in Refs.\  \cite{litvinov07,patyk08}, a distinction between the experiments in Ref.\ \cite{litvinov08} and Ref.\ \cite{kennedy75} which used stopped, presumably neutral atoms.  There is no reason {\it a priori} to expect that an electron capture decay rate oscillation would have the same amplitude in Eu compared to Pm and Pr.  Although this result would seem to disfavor the time modulation of electron capture decay, the uncertainty in the possible oscillation amplitude is large and the statistical power of this data is limited.  

\begin{figure}
\includegraphics[width=90mm,keepaspectratio=true]{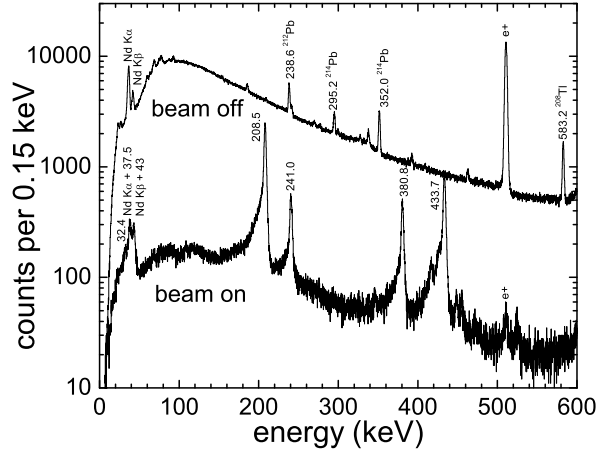}
\caption{Gamma ray spectrum measured in the clover Ge detector during and after beam bombardment.  The beam off data are the sum of all data cycles over the entire 300 second counting period, while the beam on data are the sum of all cycles during the 0.5 second bombardment.  Several common background lines are labelled in the beam off data.}
\label{fig:pmgloenergy}
\end{figure}

Despite the negative evidence from $^{142}$Eu and because of the importance of the GSI claim, we performed a test in $^{142}$Pm, one of the two isotopes reported to show a positive effect.  We studied the electron capture and beta decay rate from $^{142}$Pm, using a source of stopped ions in a foil.  A thin target (400 $\mu$g/cm$^{2}$ $^{124}$Sn backed with carbon) was bombarded with 95~MeV $^{23}$Na$^{5+}$ (average beam intensity 100 pnA) from the 88-Inch Cyclotron at Lawrence Berkeley National Laboratory.  This beam energy was selected to produce predominantly $^{142}$Pm, and calculations with PACE-2 \cite{gavron80} indicate that other $Z=61$ isotopes have production rates about a factor of 9 lower than $^{142}$Pm.  The reaction products moved through the Berkeley Gas-filled Separator (BGS) \cite{codythesis}, which separated the  $^{142}$Pm from the beam and other products by their different magnetic rigidities.  The $^{142}$Pm stopped in a 25~$\mu$m thick aluminum foil at the focal plane of the BGS.    An intrinsic germanium ``clover" detector \cite{duchene99} is located just outside a 2~mm thick aluminum vacuum window and counted the x-ray and gamma-ray emissions from the stopped $^{142}$Pm.  After a short bombardment time of 0.5~s, the primary beam was shut off, and events from the germanium detector were recorded for 300~s before repeating the beam/count cycle.  Approximately 190 bombardment cycles were performed over a 32 hour run.  With the cyclotron beam off, the germanium detector recorded roughly 300~events per second above the 20~keV threshold.  Figure \ref{fig:pmgloenergy} is the energy spectrum from the clover germanium detector, summed over all beam bombardment cycles and counting time bins.  Events during the beam bombardment are shown separately from the spectrum measured after the beam has been shut off.  Events showing energy deposition in multiple ``leaves" of the clover detector have been excluded (which reduces the Compton background under the x-ray portion of the spectrum).  K$_{\alpha 1}$ and K$_{\alpha 2}$ from the daughter $^{142}$Nd are not resolved at 37.36 and 36.85~keV, but K$_{\alpha}$ is distinguished from K$_{\beta}$ at about 42.5~keV in the ``beam off" spectrum.  The peaks at 433, 381, 241, 208, and 43~keV are cascaded from a 67~$\mu$s ($13^{-}$) state populated in the production reaction \cite{liu04}, and confirm that $^{142}$Pm was indeed produced and well-distinguished from other isobars.  These peaks disappear in the ``beam off" spectrum, confirming that the cyclotron beam was turned off, and confirming that the production time of the $^{142}$Pm was short compared to a 5 to 10 second oscillation period.

The events were sorted into time-binned energy spectra for every 0.5~s.  To measure the electron capture decays of $^{142}$Pm, we measured the rate of detected K-shell x-rays (K$_{\alpha 1}$, K$_{\alpha 2}$).  Electron capture decay strongly favors K-shell x-ray emission compared to positron decay.  The electron capture branching ratio for neutral $^{142}$Pm is 22\%, of which about 85\% is K-shell capture \cite{nucldat,bambynek77}.  The x-ray fluorescence yield for a K-shell vacancy (K$_{\alpha}$ plus  K$_{\beta}$) for Nd is 91\% \cite{krause79}.  Positron emission by $^{142}$Pm can in principle produce K-shell vacancy and hence K-shell x-rays via shakeoff and direct collision.  However, these processes strongly favor outer shell vacancy production, particularly in high Z atoms.  The K-shell vacancy probability after $\beta^{+}$ decay should be very small -- about $10^{-4}$ for $^{142}$Pm \cite{law72a,law72b,intemann83}.  A very conservative upper limit on the amount of K-shell x-rays arising from $\beta^{+}$ emission would be 1.5\%, assuming that all the Auger K electron yield from $^{142}$Pm decay \cite{nucldat} was attributable to $\beta^{+}$.  A detected K-shell x-ray therefore has a greater than 92\% probability to have been created by electron capture decay, and a less than 8\% probability to have come from positron emission.  Detection of a K-shell x-ray therefore strongly selects electron capture decays.  

\begin{figure}
\begin{center}
\includegraphics[width=90mm,keepaspectratio=true]{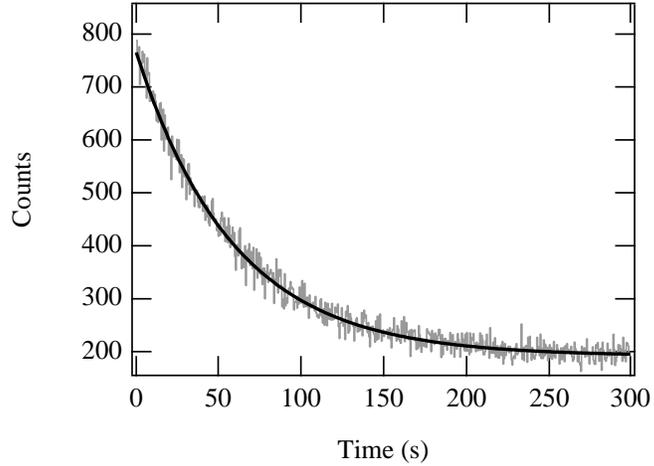}
\caption{Decay of the $^{142}$Nd K$_{\alpha}$ x-rays.  The number of counts in a window surrounding the K$_{\alpha}$ peak is plotted as a function of time after production of $^{142}$Pm. The best fit exponential decay curve is also shown.
}
\label{fig:pmKadecay}
\end{center}
\end{figure}

Figure \ref{fig:pmKadecay} shows the time decay of the Nd K$_{\alpha}$ x-rays for the entire 300~s counting time along with a best-fit exponential decay curve.  For each 0.5~s time bin after the beam bombardment, we histogram the number of counts in a 5~keV  window surrounding the K$_{\alpha}$ peak.  No oscillation is apparent in Fig.\ \ref{fig:pmKadecay}.  We performed $\chi^{2}$ minimizing fits of the decay data to the function in Eq.\ \ref{eq:oscdec} to search for a resolved oscillatory time dependence.  Fixing $A = \phi = \omega = 0$ and allowing $B$, $N_{0}$, and $T_{1/2}$ to vary for a simple exponential decay gave $\chi^{2} = 618.5$ with 594 degrees of freedom.  Allowing $B$, $N_{0}$, $T_{1/2}$, $A$, $\phi$, and $\omega$ to vary in the fits, we find a minimum $\chi^{2}/\mathrm{dof} = 614.6/591$, with $A = 0.0145(74)$, $\phi = -1.93(76)$, and an oscillation period of 3.178(36)~s.  

To evaluate the statistical significance of the fits using Eq.\ \ref{eq:oscdec} compared to a simple exponential, we performed identical fits of Monte-Carlo generated data which had similar statistical power.  The Monte-Carlo generated data contained only exponential decay with the same time-bin structure as the experimental data, with no oscillating terms.  Repeated trials of the fit procedure to Eq.\ \ref{eq:oscdec} on several hundred randomized exponential decay data sets found an average oscillation amplitude $A\neq0$ by 2.5 standard deviations.  Moreover, the $\chi^{2}$ parameter of the fits using oscillation terms improved, according to a Fisher F-test, at a confidence level of 5\%.  
Comparing the fits of the real data using a Fisher F-test, we find a confidence level of 29.5\% for the "null hypothesis" that the oscillating terms are real.  That is, the hypothesis that there {\it is} an oscillating term in the real $^{142}$Pm decay data would only be justified at a 29.5\% confidence level, compared to the 5\% confidence level generated by the fits to simulated, non-oscillatory data, and compared to the 0.6\% confidence claimed in \cite{litvinov08}).  This supports the conclusion that the fits to our data from $^{142}$Pm do not include statistically significant oscillations.  The Fisher F-test confidence levels are misleading in this instance.  

\begin{figure}
\begin{center}
\includegraphics[width=90mm,keepaspectratio=true]{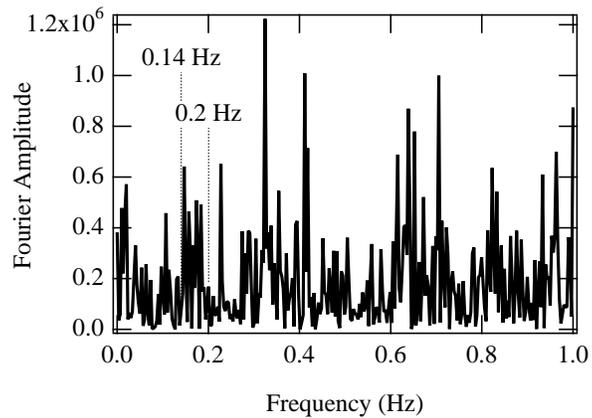}
\caption{Fourier transform of the residuals to a fit of the data shown in Fig.\ \ref{fig:pmKadecay} to a simple exponential decay.  There are no statistically resolved peaks in the data corresponding to oscillations at 7~s (0.14~Hz marker) or 5~s (0.2~Hz marker).}
\label{fig:FFTresid}
\end{center}
\end{figure}

\begin{figure}
\begin{center}
\includegraphics[width=90mm,keepaspectratio=true]{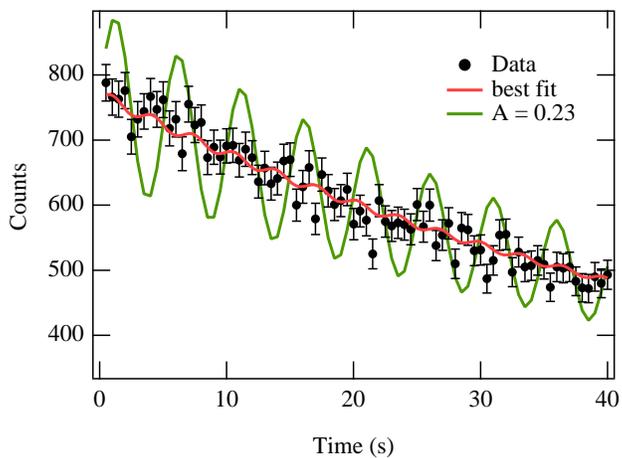}
\caption{Data from Fig.\ \ref{fig:pmKadecay} for the first 40 seconds of decay, as examined in Ref.\ \cite{litvinov08}.  The best fit to Eq.\ \ref{eq:oscdec} is shown (red), finding $A = 0.0145(74)$ and $2\pi/\omega = T_{d} = 3.178(36)$~s, as is a curve (green) calculated with an oscillation amplitude $A=0.23$ and $T_{d} = 5$~s, suggested by Ref.\ \cite{litvinov08}.}
\label{fig:pmcompare}
\end{center}
\end{figure}

Fourier transforms of the residuals to a simple exponential fit to the data are shown in Fig.\  \ref{fig:FFTresid}, and show no statistically significant peaks at 1/(5~s) (or 1/(7~s), assuming that the oscillation period does not depend on $\gamma$).  In all fit cases, we find $T_{1/2} = 40.68(53)$~s, which agrees with the accepted lifetime of $^{142}$Pm (40.5(5)~s \cite{nucldat}).  Figure \ref{fig:pmcompare} shows the first 40~seconds of our decay data, roughly the time period considered in \cite{litvinov08} for $^{142}$Pm, along with a calculated decay curve with $A = 0.23$, $T_{d} = 7.10$~s, and $\phi = -1.6$~rad found in \cite{litvinov08}.  These parameters are inconsistent with the data.  We conclude that any oscillation of 5 (or 7) seconds, if present, must have an amplitude smaller than that found in Ref.\ \cite{litvinov08} ($A = 0.23(4)$) by about a factor of 31, using the uncertainty in the oscillation amplitude fit parameter as our sensitivity.  Our beam bombardment time of 0.5 s would reduce our sensitivity to an oscillation amplitude, but a simple calculation (integrating oscillating exponential decay and production) suggests that this would reduce our sensitivity by 2\% (i.e. if $A = 0.23$, this experiment would have measured $A = 0.225$).  Our best-fit oscillation amplitude is not statistically significant, and in any case has an amplitude a factor of 16 smaller than that proposed in \cite{litvinov08}, with a much different period and phase.  

As a cross-check on our data, we measured the decay of 511~keV positron annihilation activity.  Reference \cite{litvinov08} suggests that the decay rate for electron capture events will oscillate while the $\beta^{+}$ decay rate might not, since the three-body phase space density for positron emission has a large neutrino momentum distribution, which would average out any coherence between final neutrino mass states.  Reference \cite{ivanov08b} concludes that the positron decay of hydrogen-like ions will oscillate on a time scale too fast to have been observed in our data.  The 511~keV decay data was fit to a simple exponential decay, finding a half-life of $T_{1/2} = 41.11(38)$~s with $\chi^{2}/\mathrm{dof} = 603.3$/596.  Fitting this data to Eq.\ \ref{eq:oscdec} we find a minimum $\chi^{2}/\mathrm{dof} = 597.2/593$ for $A = 0.0173(70)$, $\phi = 2.15(59)$ and period $T_{d} = 0.8129(8)$~s.  The confidence level for this fit compared to simple exponential decay is 2.8\%, suggesting that the oscillatory fit parameters improve the fit, more so than in the electron capture data set analyzed here, and slightly more so than in our Monte-Carlo simulated data which containted no oscillations. This is contrary to the expectation under the neutrino mixing hypothesis that there should be no oscillation in the positron decay activity.  This result supports the conclusion that our oscillation fit parameters in the electron capture decay data are not statistically significant.

It might be argued that our experiment using neutral atoms would be insensitive to the proposed neutrino oscillation effect, since the participation of the remaining atomic electrons could provide a decoherence of the neutrino momentum states in the larger phase space of the final atomic states after the decay.  However, the usual description of electron capture decay suggests that the neutrino and recoil nucleus momenta are determined largely by the structure of the weak interaction hamiltonian.  In the case where multiple atomic electrons are present, there can in principle be interference between contributions from different atomic electronic states.  In calculating the total decay rate, or the ratio of K-shell to L-shell electronic capture, or the ratio of electron capture to positron decay, the neutrino (and recoil nucleus) momentum is determined by an overlap integral of the atomic electrons' wave functions, summed over electron states \cite{bambynek77}.  But these calculations, particularly in high-Z nuclei such as Nd, are dominated by the K-shell contribution, and multiple electron terms represent corrections to the total decay rate of only a few percent \cite{bambynek77}.  Moreover, our experiment detected K-shell x-rays, meaning that the captured electron was indeed a K-shell electron with a similar wavefunction to the the hydrogenic ions investigated in \cite{litvinov08}.  The subsequent atomic de-excitation processes do not greatly influence the generation of the neutrino or recoil momenta.  If multiple electron effects destroy the coherence of the mixed neutrinos' momenta in the final state, this would be apparent in data from the GSI group comparing the decay time spectrum of hydrogen-like and helium-like stored ions, similar to data reported in \cite{litvinov07}.  A further desirable confirmation of the data from the GSI group would be to examine the $\beta^{+}$ decays of the hydrogen-like ions, which should show no oscillations on the timescales available for examination, according to Ref.\ \cite{ivanov08b}.

To summarize, no convincing oscillation was observed in the decay time spectrum of electron capture decays of $^{142}$Pm (or in $^{142}$Eu using published data) when dressed with their full complement of electrons and at rest in a solid metal matrix.  Any 5 second oscillation not resolved in this experiment must have an amplitude a factor of 31 times smaller than that reported in Ref.\ \cite{litvinov08}. 
The proposed oscillating decay rate could in principle be attributable to the truly two-body nature of the final state in the hydrogen-like decays observed in \cite{litvinov08}, although this would require an unconventional explanation with respect to electron capture decay in neutral atoms.  Under this hypothesis, data on the decay of helium-like stored ions would show much different oscillation behavior.  

\section*{Acknowledgments}

We appreciate the assistance of the technical staff and operators at the 88-Inch Cyclotron.  This work was supported by the Director, Office of Science, Office of Nuclear Physics, U.S.\ Department of Energy under Contract No.\ DE-AC02-05CH11231f.  Part of this work was performed under the auspices of the U.S. Department of Energy Lawrence Livermore National Laboratory under contract DE-AC52-07NA27344.

\end{document}